\documentclass[aps, prl, reprint, nofootinbib]{revtex4-2}
\newcommand{\jpsi}{J/\psi}
\newcommand{\psip}{\psi'}

\newcommand{\elp}{e^{+}}
\newcommand{\elm}{e^{-}}

\newcommand{\pip}{\pi^{+}}
\newcommand{\pim}{\pi^{-}}
\newcommand{\piz}{\pi^{0}}

\newcommand{\kp}{K^{+}}
\newcommand{\km}{K^{-}}

\newcommand{\ks}{K_{S}}

\usepackage{bm}
\usepackage{dcolumn}
\usepackage{multirow}
\usepackage{makecell}
\usepackage{diagbox}
\usepackage{comment}
\usepackage{comment}
\usepackage{indentfirst}
\usepackage{float}
\usepackage{enumitem}

\usepackage{appendix}

\usepackage[english]{babel}

\usepackage[letterpaper,top=2cm,bottom=2cm,left=3cm,right=3cm,marginparwidth=1.75cm]{geometry}

\usepackage{amsmath}
\usepackage{graphicx}
\usepackage[colorlinks=true, allcolors=blue]{hyperref}

\begin{document}

\title{Pinpointing Physical Solutions in $Y(4230)$ Decays}

\author{Jie Yuan$^{1,2}$, Yadi Wang$^1$, Kai Zhu$^2$\\
\vspace{0.2cm}
\it $^{1}$ North China Electric Power University, Beijing 102206, People's Republic of China\\
$^{2}$ Institute of High Energy Physics, Beijing 100049, People's Republic of China
}

\date{\today}
\begin{abstract}
To resolve ambiguities from multiple solutions in experimental measurements, we construct a $\chi^2$ function incorporating constraints such as isospin conservation and amplitude relations. By minimizing the global $\chi^2$, we identify physical solutions for seven hidden-charm decay channels of $Y(4230)$. Crucially, the physical solution for $Y(4230) \to \pi^{+}\pi^{-} J/\psi$ corresponds to the largest among four experimental solutions, potentially modifying inputs for theoretical calculations. Additionally, we predict $\Gamma_{ee}\mathcal{B}(Y(4230) \to \pi^{+}\pi^{-} \psip)$ based on our results, acknowledging substantial uncertainties in current measurements.

\end{abstract}

\maketitle

Since its observation, the vector resonance $Y(4230)$ (also referred to as $Y(4260)$ or $\psi(4230)$) has been regarded as a promising exotic charmonium-like candidate~\cite{BaBar:2005hhc}. Numerous studies have investigated its nature, proposing various interpretations such as a conventional charmonium state~\cite{ccbar3,ccbar6}, a hadronic molecule ($D_{1} \bar{D} + \text{c.c.}$ or $D_0 \bar{D}^* + \text{c.c.}$)~\cite{molecule,dd2,dd3,dd4,dd5,dd6,dd7,dd9,dd10,Quantum}, a tetraquark state $c q \bar{c} \bar{q}$ ($q = u,\ d,\ \text{or}\ s$)~\cite{fourquark1,fourquark2,fourquark3,fourquark4,fourquark5,fourquark8}, a $c\bar{c}$-gluon hybrid state~\cite{hybrid,hy1,hy3,hy6}, mixed charmonium-tetraquark or charmonium-molecule states~\cite{fourquark6,fourquark7,dd8}, a $\rho \chi_{c1}$ or $\omega \chi_{c1}$ molecule~\cite{Liu:2005ay,dd11}, and non-resonant explanations~\cite{ccbar4,ccbar5}. Nevertheless, a consensus regarding its exact structure has yet to be reached.

In addition to mass and width analyses, the decay channels of $Y(4230)$ provide a valuable probe into its properties. Studying these decays offers insight into the decay mechanisms and, consequently, the internal structure of the resonance. However, despite extensive experimental measurements, the presence of multiple solutions introduces significant ambiguity in the interpretation of results. This issue arises unavoidably when interference effects are incorporated, particularly when resonances are modeled with the Breit-Wigner formula~\cite{Bukin:2007kx,Zhu:2011ha,Bai:2019jrb}. When the cross-section line shape is described by coherent amplitudes, as in the expression $\left| \sum_{i} g_{i} e^{i\phi_{i}} A_{i} \right|^{2}$, multiple distinct sets of parameters $(g_{i}, \phi_{i})$ can yield identical goodness-of-fit values. As a result, identifying the physically meaningful solution based solely on experimental data remains challenging.

Recent measurements by the BESIII Collaboration of processes such as $\elp \elm \to \pi^{+} \pi^{-} J/\psi$~\cite{pipijpsi}, $\elp \elm \to \pi^{0} \pi^{0} J/\psi$~\cite{pi0pi0jpsi}, $\elp \elm \to K^{+} K^{-} J/\psi$~\cite{kkjpsi}, $\elp \elm \to K_{S} K_{S} J/\psi$~\cite{ksksjpsi}, $\elp \elm \to \eta J/\psi$~\cite{etajpsi}, $\elp \elm \to \eta' J/\psi$~\cite{eta'jpsi}, and $\elp \elm \to \pi^{+} \pi^{-} \psi'$~\cite{pipipsip} have reported the products $\Gamma_{ee}\mathcal{B}$ for the decay $Y(4230) \to e^+ e^-$, all of which exhibit multiple solutions, as summarized in Table~\ref{ta:value}. Compounding this ambiguity, some theoretical studies have misinterpreted these results. For instance, Refs.~\cite{dd6,dd7,dd9,fourquark1,hy3} erroneously equated the cross-section of $e^+ e^- \to \pi^+ \pi^- J/\psi$ at $\sqrt{s} = 4.23$~GeV with that of $e^+ e^- \to Y(4230) \to \pi^+ \pi^- J/\psi$, despite clear discrepancies evident in Table~\ref{ta:value}. Thus, identifying the physical solution among the multiple options is essential to reducing experimental uncertainties and providing reliable inputs for theoretical studies.

A previous attempt to identify the physical solutions for $Y(4230) \to \eta J/\psi$ and $Y(4230) \to \eta' J/\psi$ through $\eta$-$\eta'$ mixing analysis was made in Ref.~\cite{eta-eta'}. However, that work assumed the absence of an $s\bar{s}$ component in $Y(4230)$, considering only $c\bar{c}$ and $q\bar{q}$ constituents. In this Letter, we perform a global analysis of the existing experimental results for hidden-charm final states listed in Table~\ref{ta:value}. We aim to identify the most plausible physical solutions by incorporating constraints from symmetry conservation and other physical considerations, and simultaneously determine the fraction of the $s\bar{s}$ component. This approach will help resolve existing ambiguities in experimental data and provide a more solid foundation for subsequent theoretical investigations.

\begin{table*}[!htbp]
\setlength{\abovecaptionskip}{1pt} 
\setlength{\belowcaptionskip}{4pt}
\caption{Multiple solutions in the decays of $Y(4230)$. The products of the partial width $\Gamma_{ee}$ and branching fraction $\mathcal{B}$ (i.e., $\Gamma_{ee}\mathcal{B}$) for each decay mode are listed in units of eV. Values are taken from recent BESIII results: $\elp \elm \to \pi^{+} \pi^{-} J/\psi$~\cite{pipijpsi}, $\elp \elm \to \pi^{0} \pi^{0} J/\psi$~\cite{pi0pi0jpsi}, $\elp \elm \to K^{+} K^{-} J/\psi$~\cite{kkjpsi}, $\elp \elm \to K_{S} K_{S} J/\psi$~\cite{ksksjpsi}, $\elp \elm \to \eta J/\psi$~\cite{etajpsi}, $\elp \elm \to \eta' J/\psi$~\cite{eta'jpsi}, and $\elp \elm \to \pi^{+} \pi^{-} \psi'$~\cite{pipipsip}. The last column lists the factors, defined as the inverse of the product of the $Y(4230)$ mass squared and its total width, with units of $1/\mathrm{GeV}^3$, as employed in various experimental analyses.  Uncertainties include both statistical and systematic contributions.}
\centering
\setlength{\tabcolsep}{5pt}
\renewcommand{\arraystretch}{1.5}
\label{ta:value}
\begin{tabular}{cccccc} 
\hline \hline 
Mode & Solution I & Solution II & Solution III & Solution IV & $\frac{1}{M^2 \Gamma_{tot}}$  \\ \hline
$\pip\pim\jpsi$   & $1.7 \pm 0.2$  & $8.2 \pm 0.9$  & $3.0 \pm 0.5$  & $14.6 \pm 1.2$  &$1.3\pm 0.1$ \\
$\piz\piz\jpsi$ & $0.99 \pm 0.17$ & $4.13 \pm 0.28$ & $1.38 \pm 0.30$ & $5.72 \pm 1.57$ & $1.2\pm 0.1 $\\     
$\kp \km\jpsi$   & $0.42 \pm 0.15$ & $0.29 \pm 0.10$ & -- & -- & $0.8 \pm 0.3$\\
$\ks\ks\jpsi$   & $0.13 \pm 0.05$ & $0.14 \pm 0.06$ & $0.18 \pm 0.07$ & $0.20 \pm 0.08$ & $0.8\pm 0.4$\\
$\eta\jpsi$     & $4.0 \pm 0.5$   & $5.5 \pm 0.7$   & $8.7 \pm 1.0$   & $11.9 \pm 1.1$ & $0.7\pm 0.1$\\        
$\eta'\jpsi$    & $0.06 \pm 0.03$ & $1.38 \pm 0.11$ & -- & -- &$1.0\pm 0.3$\\
$\pip\pim\psip$ & $1.6 \pm 1.3$   & $1.8 \pm 1.4$   & -- & -- &$0.7 \pm 0.2$\\ \hline \hline
\end{tabular}
\end{table*}

To identify the physical solutions among the multiple candidates, we construct a $\chi^2$ function comprising terms derived from symmetry constraints and amplitude ratios between decay channels. All possible combinations of solutions across the decay channels are evaluated based on the $\chi^2$ value, and the combination yielding the global minimum is selected as the set of physical solutions. Below, we detail the $\chi^2$ methodology and its application to seven decay channels.

First, isospin symmetry is assumed to be well conserved in the cross-sections. Consequently, the branching fractions are expected to satisfy the ratios $\mathcal{B}(Y(4230)\to\pi^{+}\pi^{-}J/\psi) : \mathcal{B}(Y(4230)\to\pi^{0}\pi^{0}J/\psi) = 2 : 1$ and $\mathcal{B}(Y(4230)\to K^{+}K^{-}J/\psi) : \mathcal{B}(Y(4230)\to K_{S}K_{S}J/\psi) = 2 : 1$. Thus, the following two terms are included in the $\chi^2$ function:
\begin{equation}
\label{eq:iso1}
\frac{(\mathcal{B}_{\pi^{+}\pi^{-}J/\psi} - 2\mathcal{B}_{\pi^{0}\pi^{0}J/\psi})^{2}}{\Delta_{\mathcal{B}_{\pi^{+}\pi^{-}J/\psi}}^{2} + (2\Delta_{\mathcal{B}_{\pi^{0}\pi^{0}J/\psi}})^{2}}    
\end{equation}
and
\begin{equation}
\label{eq:iso2}
   \frac{(\mathcal{B}_{K^{+}K^{-}J/\psi} - 2\mathcal{B}_{K_{S}K_{S}J/\psi})^{2}}{\Delta_{\mathcal{B}_{K^{+}K^{-}J/\psi}}^{2} + (2\Delta_{\mathcal{B}_{K_{S}K_{S}J/\psi}})^{2}}\ ,
\end{equation}
where $\Delta_{\mathcal{B}_X}$ indicates corresponding experimental uncertainty quoted from published papers.

We note that various experimental analyses have employed different values for the mass and total width of $Y(4230)$. To account for this variation, we introduce a correction factor defined as the inverse of the product of the $Y(4230)$ mass squared and its total width. These factors are listed in the final column of Table~\ref{ta:value}, and all branching fractions presented herein have been correspondingly corrected.

According to Ref.~\cite{explanation}, if the possible meson loop effect is ignored, the ratio between the branching fractions of $Y(4230) \to \pi^{+}\pi^{-}\psip$ and $Y(4230) \to \pi^{+}\pi^{-}J/\psi$ is given by
\[
  \frac{\mathcal{B}_{\pi^{+}\pi^{-}\psip}}{\mathcal{B}_{\pi^{+}\pi^{-}J/\psi}}  
               \approx \frac{\Omega_{\pi^{+}\pi^{-}\psip}}{\Omega_{\pi^{+}\pi^{-}J/\psi}} \frac{|\psi'(0)|^2}{|\psi(0)|^2},
\]
where $\Omega_{\pi^{+}\pi^{-}\psip}$ ($\Omega_{\pi^{+}\pi^{-}J/\psi}$) is the phase space for $Y(4230) \to \pi^{+}\pi^{-}\psip$ ($Y(4230) \to \pi^{+}\pi^{-}J/\psi$), and $\psi'(0)$ ($\psi(0)$) is the $\psip$ ($J/\psi$) radial wave function at the origin in the quark model. 
The wave function ratio is directly related to the leptonic decay widths through the well-established relation:
\[
\frac{\Gamma(\psip \to e^+e^-)}{\Gamma(J/\psi \to e^+e^-)} = \frac{M_{J/\psi}^{2}}{M_{\psip}^{2}} \frac{|\psi'(0)|^2}{|\psi(0)|^2},
\]
where $M_{\psip}$ and $M_{J/\psi}$ are the masses of $\psip$ and $J/\psi$, respectively. Using the measured leptonic widths from the PDG~\cite{pdg}, we determine the wave function ratio $|\psi'(0)|^2/|\psi(0)|^2$. This allows us to calculate the expected branching fraction ratio $\mathcal{B}(Y(4230) \to\pi^{+}\pi^{-}\psip) : \mathcal{B}(Y(4230) \to\pi^{+}\pi^{-}\jpsi) \approx 0.04 : 1$. The corresponding term in the $\chi^2$ is
\begin{equation}
\label{eq:psip}
    \frac{(\mathcal{B}_{\pi^{+}\pi^{-}\psip} - 0.04\mathcal{B}_{\pi^{+}\pi^{-}J/\psi})^{2}}{\Delta_{\mathcal{B}_{\pi^{+}\pi^{-}\psip}}^{2} + (0.04\Delta_{\mathcal{B}_{\pi^{+}\pi^{-}J/\psi}})^{2}}.
\end{equation} 

From the analysis of $\eta$-$\eta'$ mixing in Ref.~\cite{eta-eta'}, the ratio of the decay amplitudes is
\[
\left| \frac{M_{\eta'}}{M_{\eta}} \right| = \left| \frac{\cos(\theta_0 - \theta) + \delta \sin(\theta_0 - \theta)}{\sin(\theta_0 - \theta) - \delta \cos(\theta_0 - \theta)} \right|,
\]
where $\theta_0 = \arctan(1/\sqrt{2}) \approx 35.3^\circ$ and $\theta = -14.8^{\circ} \pm 0.5^{\circ}$ is the empirical mixing angle~\cite{eta-eta'}. The branching fraction ratio is then
\[
   r_\eta(\delta) := \frac{\mathcal{B}(Y(4230)\to\eta'J/\psi)}{\mathcal{B}(Y(4230)\to\eta J/\psi)} = \left| \frac{M_{\eta'}}{M_{\eta}} \right|^{2} \frac{\Omega_{\eta'}}{\Omega_{\eta}},
\]
with $\Omega_{\eta'}/\Omega_{\eta} = 0.22$. The corresponding $\chi^2$ term includes the theoretical uncertainty $\Delta^\eta_{\text{theo}}$ associated with $r_\eta(\delta)$:
\begin{equation}
 \frac{\left(\mathcal{B}_{\eta' J/\psi} - r_\eta(\delta) \mathcal{B}_{\eta J/\psi}\right)^{2}}{\Delta_{\mathcal{B}_{\eta' J/\psi}}^{2} + \left(r_\eta(\delta)\Delta_{\mathcal{B}_{\eta J/\psi}}\right)^2 + (\Delta^\eta_{\text{theo}})^2 },
\end{equation}
where $\Delta^\eta_{\text{theo}} = \Delta_{r_\eta(\delta)} \mathcal{B}_{\eta J/\psi}$ quantifies the uncertainty in the theoretical calculation of $r_\eta(\delta)$.

For the ratio of $\mathcal{B}(Y(4230) \to K^{+}K^{-}J/\psi)$ over $\mathcal{B}(Y(4230) \to \pi^{+}\pi^{-}J/\psi)$, we consider the production mechanisms where $\pi^{+}\pi^{-}$ and $K^{+}K^{-}$ final states predominantly originate from light quark ($q\bar{q}$, $q=u/d$) and strange quark ($s\bar{s}$) components, respectively. Following Ref.~\cite{hiddencharm}, we account for mixing between these components through intermediate resonances like $f_0(980)$. This mixing transforms the initial quark components according to:
\[
\begin{pmatrix}
\vert q\bar{q} \rangle\langle q\bar{q} \vert & g \vert q\bar{q} \rangle\langle s\bar{s} \vert \\
g \vert s\bar{s} \rangle\langle q\bar{q} \vert & \vert s\bar{s} \rangle\langle s\bar{s} \vert
\end{pmatrix}
\begin{pmatrix}
\vert q\bar{q} \rangle \\
\delta\vert s\bar{s} \rangle
\end{pmatrix}
=
\begin{pmatrix}
\left( 1+ g\delta \right)\vert q\bar{q} \rangle \\
\left( g + \delta \right)\vert s\bar{s} \rangle
\end{pmatrix},
\]
assuming orthogonal states $\vert q\bar{q} \rangle$ and $\vert s\bar{s} \rangle$. Here, $g$ is the mixing parameter determined to be $g^2 = 4.16 \pm 0.44$ from Ref.~\cite{hiddencharm}, which originally assumed no intrinsic $s\bar{s}$ component in $Y(4230)$. Among the three methods in Ref.~\cite{hiddencharm} (narrow-width approximation, phase-space improved approximation, and direct momentum integration), we adopt the last approach. The resulting branching fraction ratio is:
\[
r_k(\delta) := \frac{\mathcal{B}_{K^{+}K^{-}J/\psi}}{\mathcal{B}_{\pi^{+}\pi^{-}J/\psi}} = \left|\frac{g+\delta}{1+g\delta}\right|^2 \frac{\Omega_{\kp\km\jpsi}}{\Omega_{\pip\pim\jpsi}}.
\]
The corresponding $\chi^2$ term incorporates both experimental and theoretical uncertainties:
\begin{equation}
\label{eq:kkpipi}
 \frac{\left(\mathcal{B}_{K^{+}K^{-}J/\psi} - r_k(\delta) \mathcal{B}_{\pi^{+}\pi^{-}J/\psi}\right)^{2}}{\Delta_{\mathcal{B}_{K^{+}K^{-}J/\psi}}^{2} + \left(r_k(\delta) \Delta_{\mathcal{B}_{\pi^{+}\pi^{-}J/\psi}}\right)^{2} + \left( \Delta^k_{\text{theo}} \right)^{2} },
 \end{equation}
where $\Delta^k_{\text{theo}} = \Delta_{r_k(\delta)} \mathcal{B}_{\pi^{+}\pi^{-}J/\psi}$ quantifies the theoretical uncertainty in calculating $r_k(\delta)$, primarily from the mixing parameter $g$ and the orthogonal state assumption.

\begin{figure}[htbp]
    \centering
    \includegraphics[width=0.48\textwidth]{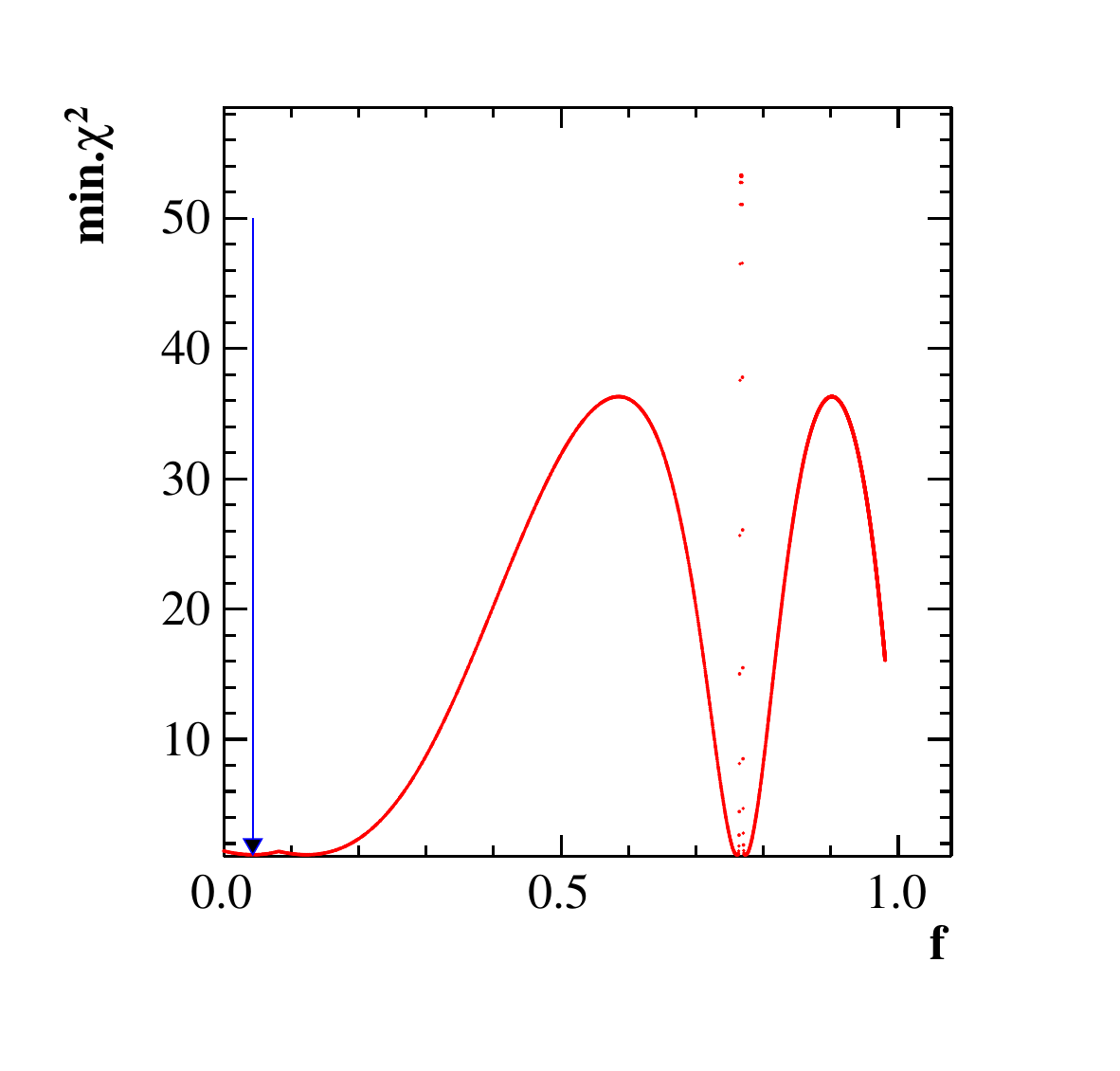}
    \caption{Minimum $\chi^2$ as a function of the strange quark fraction $f$. The blue arrow indicates the global minimum.}
    \label{fig:delta}
\end{figure}

The total $\chi^2$ function is the sum of all terms from Eq.~\eqref{eq:iso1} to \eqref{eq:kkpipi}. We introduce the fraction $f = \delta/\sqrt{1+\delta^2}$ representing the proportion of $s\bar{s}$ component. The dependence of the minimum $\chi^2$ on $f$ is shown in Fig.~\ref{fig:delta}. The global minimum occurs at $f = 0.043$ with $\chi^2 = 1.17$, and the corresponding solution set is adopted as the physical solutions, as presented in Table~\ref{tab:result}.

A particularly significant outcome is the selection of the largest solution for $Y(4230) \to \pi^+ \pi^- J/\psi$ as the physical solution. This solution differs from the bare cross section $e^+ e^- \to \pi^+ \pi^- J/\psi$ at $\sqrt{s} = 4.23$ GeV, which has been directly quoted by Refs.~\cite{dd6,dd7,dd9,fourquark1,hy3}. We recommend that these theoretical calculations be updated using the $Y(4230) \to \pi^+ \pi^- J/\psi$ cross section determined in our analysis. Experimentally, BESIII recently conducted a partial wave analysis of $Y(4230) \to \pi^+ \pi^- J/\psi$~\cite{BESIIIPWA}. For each intermediate process—$e^+ e^- \to \pi^{\pm} Z_{c}(3900)^{\mp} \to \pi^+ \pi^- J/\psi$ and $e^+ e^- \to f_{0}(980) J/\psi \to \pi^+ \pi^- J/\psi$—three solutions were obtained. Our results indicate that the solutions with the largest cross-sections should be considered physical.

From Fig.~\ref{fig:delta}, the minimum $\chi^2$ assuming substantial strange quark component ($f=0.762$) is only marginally larger than the global minimum at $f = 0.043$. However, as shown in Table~\ref{tab:result}, the physical solutions for $f=0.043$ and $f=0.762$ exhibit no substantial differences except the $\eta' J/\psi$ channel. The peak observed in Fig.~\ref{fig:delta}, on $f=0.767$, originates from a pole in the amplitude ratio for $Y(4230) \to \eta' J/\psi$ versus $\eta J/\psi$.

Table~\ref{ta:part} presents the minimum $\chi^2$ for each constraint term and the corresponding decay ratios based on the physical solutions. All terms yield $\chi^2$ values significantly below 1. This observation may result from either the inclusion of systematic uncertainties in experimental measurements or overestimated uncertainties. The relatively large $\chi^2$ value for $\mathcal{B}(Y(4230) \to \pi^+ \pi^- \psip)/\mathcal{B}(Y(4230) \to \pi^+ \pi^- J/\psi)$ indicates possible deviation in the mean value of $\mathcal{B}(Y(4230) \to \pi^+ \pi^- \psip)$, despite its considerable uncertainty. Based on our analysis and the physical solution for $\mathcal{B}(Y(4230) \to \pi^+ \pi^- J/\psi)$, we predict $\Gamma_{ee}\mathcal{B}(Y(4230)\to \pi^+ \pi^- \psip) = 0.3 \pm 0.1 ~\text{eV}$, a value that can be tested in future experimental measurements.

It should be noted that several constraints adopted in this study do not incorporate long-range effects, which are commonly treated as meson loop contributions. Consequently, caution is advised when applying these results to theoretical frameworks where meson loop effects play a substantial role.

\begin{table}[htbp]
\centering
\caption{The physical solutions of the $Y(4230)$ decays corresponding to the global minimum $\chi^{2}$. The products of the partial width $\Gamma_{ee}$ and branching fraction $\mathcal{B}$ (i.e., $\Gamma_{ee}\mathcal{B}$) for each decay mode are listed in units of eV. }
\begin{tabular}{ccc}\hline \hline
$f$ & 0.043  &  0.762\\ \hline 
$ \pip\pim\jpsi$       & 14.6 ± 1.2    & 14.6 ± 1.2     \\ 
$\piz\piz\jpsi$    & 5.72± 1.57  & 5.72± 1.57  \\
$\kp \km\jpsi$  & 0.29$\pm$0.10   & 0.29$\pm$0.10 \\
$\ks\ks\jpsi$  & 0.14$\pm$0.06   & 0.14$\pm$0.06  \\
$\eta\jpsi$    & $5.5 \pm 0.7$      & $4.0 \pm 0.5$      \\
$\eta'\jpsi$        & 1.38 ± 0.11 & 0.06 ± 0.03    \\
$\pip\pim\psip$       & 1.6 ± 1.3  & 1.6 ± 1.3 \\  \hline

$\chi^2$  & 1.17  &1.18 \\ \hline \hline
\end{tabular}
\label{tab:result}
\end{table}

\begin{table}[htbp]
\centering
\caption{The $\chi^{2}$ contributions of each constraint term and decay ratio based on the physical solutions. Under both scenarios with the proportion of $s\bar{s}$ component $f=0.043$ and $f=0.762$ the results are presented.}
\begin{tabular}{ccccc} \hline \hline
          & \multicolumn{2}{c}{$f$=0.043} &\multicolumn{2}{c}{$f$=0.762}\\ \hline
Constraint term         & $\chi^{2}$            &  ratio    &$\chi^{2}$    &  ratio \\ \hline
$B(\pip\pim\jpsi)/B(\piz\piz\jpsi)$       & 0.22  & 2.30     & 0.22 & 2.30\\ 
$B(\kp\km\jpsi)/B(\ks\ks\jpsi)$           & 0.01  & 2.10     & 0.01 & 2.20\\ 
$B(\pip\pim\psip)/B(\pip\pim\jpsi)$       & 0.92  & 0.21     & 0.92 & 0.21  \\ 
$B(\eta'\jpsi)/B(\eta\jpsi) $             & 0.01  & 0.18     & 0.02 & 0.01 \\ 
$B(\kp\km\jpsi)/B(\pip\pim\jpsi)$         & 0.01  & 0.03     & 0.01 &0.03  \\ \hline
tot   & 1.17 &  &1.18  \\ \hline \hline                     
\end{tabular}
\label{ta:part}
\end{table}

\section{Acknowledgement}
National Natural Science Foundation of China (NSFC) under Contracts No. 12375083 and No. 12105100; Beijing Natural Science Foundation (BJNSF) under Contract No. JQ22002.

\bibliographystyle{apsrev4-2}
\bibliography{reference}

\begin{thebibliography}{45}%
\makeatletter
\providecommand \@ifxundefined [1]{%
 \@ifx{#1\undefined}
}%
\providecommand \@ifnum [1]{%
 \ifnum #1\expandafter \@firstoftwo
 \else \expandafter \@secondoftwo
 \fi
}%
\providecommand \@ifx [1]{%
 \ifx #1\expandafter \@firstoftwo
 \else \expandafter \@secondoftwo
 \fi
}%
\providecommand \natexlab [1]{#1}%
\providecommand \enquote  [1]{``#1''}%
\providecommand \bibnamefont  [1]{#1}%
\providecommand \bibfnamefont [1]{#1}%
\providecommand \citenamefont [1]{#1}%
\providecommand \href@noop [0]{\@secondoftwo}%
\providecommand \href [0]{\begingroup \@sanitize@url \@href}%
\providecommand \@href[1]{\@@startlink{#1}\@@href}%
\providecommand \@@href[1]{\endgroup#1\@@endlink}%
\providecommand \@sanitize@url [0]{\catcode `\\12\catcode `\$12\catcode
  `\&12\catcode `\#12\catcode `\^12\catcode `\_12\catcode `\%12\relax}%
\providecommand \@@startlink[1]{}%
\providecommand \@@endlink[0]{}%
\providecommand \url  [0]{\begingroup\@sanitize@url \@url }%
\providecommand \@url [1]{\endgroup\@href {#1}{\urlprefix }}%
\providecommand \urlprefix  [0]{URL }%
\providecommand \Eprint [0]{\href }%
\providecommand \doibase [0]{https://doi.org/}%
\providecommand \selectlanguage [0]{\@gobble}%
\providecommand \bibinfo  [0]{\@secondoftwo}%
\providecommand \bibfield  [0]{\@secondoftwo}%
\providecommand \translation [1]{[#1]}%
\providecommand \BibitemOpen [0]{}%
\providecommand \bibitemStop [0]{}%
\providecommand \bibitemNoStop [0]{.\EOS\space}%
\providecommand \EOS [0]{\spacefactor3000\relax}%
\providecommand \BibitemShut  [1]{\csname bibitem#1\endcsname}%
\let\auto@bib@innerbib\@empty
\bibitem [{\citenamefont {Aubert}\ \emph {et~al.}(2005)\citenamefont {Aubert}
  \emph {et~al.}}]{BaBar:2005hhc}%
  \BibitemOpen
  \bibfield  {author} {\bibinfo {author} {\bibfnamefont {B.}~\bibnamefont
  {Aubert}} \emph {et~al.} (\bibinfo {collaboration} {BaBar}),\ }\href
  {https://doi.org/10.1103/PhysRevLett.95.142001} {\bibfield  {journal}
  {\bibinfo  {journal} {Phys. Rev. Lett.}\ }\textbf {\bibinfo {volume} {95}},\
  \bibinfo {pages} {142001} (\bibinfo {year} {2005})},\ \Eprint
  {https://arxiv.org/abs/hep-ex/0506081} {arXiv:hep-ex/0506081} \BibitemShut
  {NoStop}%
\bibitem [{\citenamefont {Bokade}\ and\ \citenamefont
  {Bhaghyesh}(2025)}]{ccbar3}%
  \BibitemOpen
  \bibfield  {author} {\bibinfo {author} {\bibfnamefont {C.~A.}\ \bibnamefont
  {Bokade}}\ and\ \bibinfo {author} {\bibnamefont {Bhaghyesh}},\ }\href
  {https://doi.org/10.1103/PhysRevD.111.014030} {\bibfield  {journal} {\bibinfo
   {journal} {Phys. Rev. D}\ }\textbf {\bibinfo {volume} {111}},\ \bibinfo
  {pages} {014030} (\bibinfo {year} {2025})}\BibitemShut {NoStop}%
\bibitem [{\citenamefont {Llanes-Estrada}(2005)}]{ccbar6}%
  \BibitemOpen
  \bibfield  {author} {\bibinfo {author} {\bibfnamefont {F.~J.}\ \bibnamefont
  {Llanes-Estrada}},\ }\href {https://doi.org/10.1103/PhysRevD.72.031503}
  {\bibfield  {journal} {\bibinfo  {journal} {Phys. Rev. D}\ }\textbf {\bibinfo
  {volume} {72}},\ \bibinfo {pages} {031503} (\bibinfo {year}
  {2005})}\BibitemShut {NoStop}%
\bibitem [{\citenamefont {Ding}(2009)}]{molecule}%
  \BibitemOpen
  \bibfield  {author} {\bibinfo {author} {\bibfnamefont {G.-J.}\ \bibnamefont
  {Ding}},\ }\href {https://doi.org/10.1103/PhysRevD.79.014001} {\bibfield
  {journal} {\bibinfo  {journal} {Phys. Rev. D}\ }\textbf {\bibinfo {volume}
  {79}},\ \bibinfo {pages} {014001} (\bibinfo {year} {2009})},\ \Eprint
  {https://arxiv.org/abs/0809.4818} {arXiv:0809.4818 [hep-ph]} \BibitemShut
  {NoStop}%
\bibitem [{\citenamefont {Wang}\ \emph {et~al.}(2013)\citenamefont {Wang},
  \citenamefont {Hanhart},\ and\ \citenamefont {Zhao}}]{dd2}%
  \BibitemOpen
  \bibfield  {author} {\bibinfo {author} {\bibfnamefont {Q.}~\bibnamefont
  {Wang}}, \bibinfo {author} {\bibfnamefont {C.}~\bibnamefont {Hanhart}},\ and\
  \bibinfo {author} {\bibfnamefont {Q.}~\bibnamefont {Zhao}},\ }\href
  {https://doi.org/10.1103/PhysRevLett.111.132003} {\bibfield  {journal}
  {\bibinfo  {journal} {Phys. Rev. Lett.}\ }\textbf {\bibinfo {volume} {111}},\
  \bibinfo {pages} {132003} (\bibinfo {year} {2013})}\BibitemShut {NoStop}%
\bibitem [{\citenamefont {von Detten}\ \emph {et~al.}(2024)\citenamefont {von
  Detten}, \citenamefont {Baru}, \citenamefont {Hanhart}, \citenamefont {Wang},
  \citenamefont {Winney},\ and\ \citenamefont {Zhao}}]{dd3}%
  \BibitemOpen
  \bibfield  {author} {\bibinfo {author} {\bibfnamefont {L.}~\bibnamefont {von
  Detten}}, \bibinfo {author} {\bibfnamefont {V.}~\bibnamefont {Baru}},
  \bibinfo {author} {\bibfnamefont {C.}~\bibnamefont {Hanhart}}, \bibinfo
  {author} {\bibfnamefont {Q.}~\bibnamefont {Wang}}, \bibinfo {author}
  {\bibfnamefont {D.}~\bibnamefont {Winney}},\ and\ \bibinfo {author}
  {\bibfnamefont {Q.}~\bibnamefont {Zhao}},\ }\href
  {https://doi.org/10.1103/PhysRevD.109.116002} {\bibfield  {journal} {\bibinfo
   {journal} {Phys. Rev. D}\ }\textbf {\bibinfo {volume} {109}},\ \bibinfo
  {pages} {116002} (\bibinfo {year} {2024})}\BibitemShut {NoStop}%
\bibitem [{\citenamefont {Cleven}\ \emph {et~al.}(2014)\citenamefont {Cleven},
  \citenamefont {Wang}, \citenamefont {Guo}, \citenamefont {Hanhart},
  \citenamefont {Mei\ss{}ner},\ and\ \citenamefont {Zhao}}]{dd4}%
  \BibitemOpen
  \bibfield  {author} {\bibinfo {author} {\bibfnamefont {M.}~\bibnamefont
  {Cleven}}, \bibinfo {author} {\bibfnamefont {Q.}~\bibnamefont {Wang}},
  \bibinfo {author} {\bibfnamefont {F.-K.}\ \bibnamefont {Guo}}, \bibinfo
  {author} {\bibfnamefont {C.}~\bibnamefont {Hanhart}}, \bibinfo {author}
  {\bibfnamefont {U.-G.}\ \bibnamefont {Mei\ss{}ner}},\ and\ \bibinfo {author}
  {\bibfnamefont {Q.}~\bibnamefont {Zhao}},\ }\href
  {https://doi.org/10.1103/PhysRevD.90.074039} {\bibfield  {journal} {\bibinfo
  {journal} {Phys. Rev. D}\ }\textbf {\bibinfo {volume} {90}},\ \bibinfo
  {pages} {074039} (\bibinfo {year} {2014})}\BibitemShut {NoStop}%
\bibitem [{\citenamefont {Li}\ and\ \citenamefont {Liu}(2013)}]{dd5}%
  \BibitemOpen
  \bibfield  {author} {\bibinfo {author} {\bibfnamefont {G.}~\bibnamefont
  {Li}}\ and\ \bibinfo {author} {\bibfnamefont {X.-H.}\ \bibnamefont {Liu}},\
  }\href {https://doi.org/10.1103/PhysRevD.88.094008} {\bibfield  {journal}
  {\bibinfo  {journal} {Phys. Rev. D}\ }\textbf {\bibinfo {volume} {88}},\
  \bibinfo {pages} {094008} (\bibinfo {year} {2013})}\BibitemShut {NoStop}%
\bibitem [{\citenamefont {Chen}\ \emph {et~al.}(2019)\citenamefont {Chen},
  \citenamefont {Dai}, \citenamefont {Guo},\ and\ \citenamefont {Kubis}}]{dd6}%
  \BibitemOpen
  \bibfield  {author} {\bibinfo {author} {\bibfnamefont {Y.-H.}\ \bibnamefont
  {Chen}}, \bibinfo {author} {\bibfnamefont {L.-Y.}\ \bibnamefont {Dai}},
  \bibinfo {author} {\bibfnamefont {F.-K.}\ \bibnamefont {Guo}},\ and\ \bibinfo
  {author} {\bibfnamefont {B.}~\bibnamefont {Kubis}},\ }\href
  {https://doi.org/10.1103/PhysRevD.99.074016} {\bibfield  {journal} {\bibinfo
  {journal} {Phys. Rev. D}\ }\textbf {\bibinfo {volume} {99}},\ \bibinfo
  {pages} {074016} (\bibinfo {year} {2019})}\BibitemShut {NoStop}%
\bibitem [{\citenamefont {Wu}\ \emph {et~al.}(2014)\citenamefont {Wu},
  \citenamefont {Hanhart}, \citenamefont {Wang},\ and\ \citenamefont
  {Zhao}}]{dd7}%
  \BibitemOpen
  \bibfield  {author} {\bibinfo {author} {\bibfnamefont {X.-G.}\ \bibnamefont
  {Wu}}, \bibinfo {author} {\bibfnamefont {C.}~\bibnamefont {Hanhart}},
  \bibinfo {author} {\bibfnamefont {Q.}~\bibnamefont {Wang}},\ and\ \bibinfo
  {author} {\bibfnamefont {Q.}~\bibnamefont {Zhao}},\ }\href
  {https://doi.org/10.1103/PhysRevD.89.054038} {\bibfield  {journal} {\bibinfo
  {journal} {Phys. Rev. D}\ }\textbf {\bibinfo {volume} {89}},\ \bibinfo
  {pages} {054038} (\bibinfo {year} {2014})}\BibitemShut {NoStop}%
\bibitem [{\citenamefont {Dong}\ \emph
  {et~al.}(2014{\natexlab{a}})\citenamefont {Dong}, \citenamefont {Faessler},
  \citenamefont {Gutsche},\ and\ \citenamefont {Lyubovitskij}}]{dd9}%
  \BibitemOpen
  \bibfield  {author} {\bibinfo {author} {\bibfnamefont {Y.}~\bibnamefont
  {Dong}}, \bibinfo {author} {\bibfnamefont {A.}~\bibnamefont {Faessler}},
  \bibinfo {author} {\bibfnamefont {T.}~\bibnamefont {Gutsche}},\ and\ \bibinfo
  {author} {\bibfnamefont {V.~E.}\ \bibnamefont {Lyubovitskij}},\ }\href
  {https://doi.org/10.1103/PhysRevD.90.074032} {\bibfield  {journal} {\bibinfo
  {journal} {Phys. Rev. D}\ }\textbf {\bibinfo {volume} {90}},\ \bibinfo
  {pages} {074032} (\bibinfo {year} {2014}{\natexlab{a}})}\BibitemShut
  {NoStop}%
\bibitem [{\citenamefont {Dong}\ \emph
  {et~al.}(2014{\natexlab{b}})\citenamefont {Dong}, \citenamefont {Faessler},
  \citenamefont {Gutsche},\ and\ \citenamefont {Lyubovitskij}}]{dd10}%
  \BibitemOpen
  \bibfield  {author} {\bibinfo {author} {\bibfnamefont {Y.}~\bibnamefont
  {Dong}}, \bibinfo {author} {\bibfnamefont {A.}~\bibnamefont {Faessler}},
  \bibinfo {author} {\bibfnamefont {T.}~\bibnamefont {Gutsche}},\ and\ \bibinfo
  {author} {\bibfnamefont {V.~E.}\ \bibnamefont {Lyubovitskij}},\ }\href
  {https://doi.org/10.1103/PhysRevD.89.034018} {\bibfield  {journal} {\bibinfo
  {journal} {Phys. Rev. D}\ }\textbf {\bibinfo {volume} {89}},\ \bibinfo
  {pages} {034018} (\bibinfo {year} {2014}{\natexlab{b}})}\BibitemShut
  {NoStop}%
\bibitem [{\citenamefont {Ji}\ \emph {et~al.}(2022)\citenamefont {Ji},
  \citenamefont {Dong}, \citenamefont {Guo},\ and\ \citenamefont
  {Zou}}]{Quantum}%
  \BibitemOpen
  \bibfield  {author} {\bibinfo {author} {\bibfnamefont {T.}~\bibnamefont
  {Ji}}, \bibinfo {author} {\bibfnamefont {X.-K.}\ \bibnamefont {Dong}},
  \bibinfo {author} {\bibfnamefont {F.-K.}\ \bibnamefont {Guo}},\ and\ \bibinfo
  {author} {\bibfnamefont {B.-S.}\ \bibnamefont {Zou}},\ }\href
  {https://doi.org/10.1103/PhysRevLett.129.102002} {\bibfield  {journal}
  {\bibinfo  {journal} {Phys. Rev. Lett.}\ }\textbf {\bibinfo {volume} {129}},\
  \bibinfo {pages} {102002} (\bibinfo {year} {2022})}\BibitemShut {NoStop}%
\bibitem [{\citenamefont {Maiani}\ \emph {et~al.}(2005)\citenamefont {Maiani},
  \citenamefont {Piccinini}, \citenamefont {Polosa},\ and\ \citenamefont
  {Riquer}}]{fourquark1}%
  \BibitemOpen
  \bibfield  {author} {\bibinfo {author} {\bibfnamefont {L.}~\bibnamefont
  {Maiani}}, \bibinfo {author} {\bibfnamefont {F.}~\bibnamefont {Piccinini}},
  \bibinfo {author} {\bibfnamefont {A.~D.}\ \bibnamefont {Polosa}},\ and\
  \bibinfo {author} {\bibfnamefont {V.}~\bibnamefont {Riquer}},\ }\href
  {https://doi.org/10.1103/PhysRevD.72.031502} {\bibfield  {journal} {\bibinfo
  {journal} {Phys. Rev. D}\ }\textbf {\bibinfo {volume} {72}},\ \bibinfo
  {pages} {031502} (\bibinfo {year} {2005})}\BibitemShut {NoStop}%
\bibitem [{\citenamefont {Ping}\ \emph {et~al.}(2015)\citenamefont {Ping},
  \citenamefont {Cheng-Rong},\ and\ \citenamefont {Jia-Lun}}]{fourquark2}%
  \BibitemOpen
  \bibfield  {author} {\bibinfo {author} {\bibfnamefont {Z.}~\bibnamefont
  {Ping}}, \bibinfo {author} {\bibfnamefont {D.}~\bibnamefont {Cheng-Rong}},\
  and\ \bibinfo {author} {\bibfnamefont {P.}~\bibnamefont {Jia-Lun}},\ }\href
  {https://doi.org/10.1088/0256-307X/32/10/101201} {\bibfield  {journal}
  {\bibinfo  {journal} {Chin. Phys. Lett.}\ }\textbf {\bibinfo {volume} {32}},\
  \bibinfo {pages} {101201} (\bibinfo {year} {2015})}\BibitemShut {NoStop}%
\bibitem [{\citenamefont {Wang}(2018{\natexlab{a}})}]{fourquark3}%
  \BibitemOpen
  \bibfield  {author} {\bibinfo {author} {\bibfnamefont {Z.-G.}\ \bibnamefont
  {Wang}},\ }\bibfield  {journal} {\bibinfo  {journal} {The European Physical
  Journal C}\ }\textbf {\bibinfo {volume} {78}},\ \href
  {https://doi.org/10.1140/epjc/s10052-018-6417-5}
  {10.1140/epjc/s10052-018-6417-5} (\bibinfo {year}
  {2018}{\natexlab{a}})\BibitemShut {NoStop}%
\bibitem [{\citenamefont {Zhang}\ and\ \citenamefont
  {Huang}(2011)}]{fourquark4}%
  \BibitemOpen
  \bibfield  {author} {\bibinfo {author} {\bibfnamefont {J.-R.}\ \bibnamefont
  {Zhang}}\ and\ \bibinfo {author} {\bibfnamefont {M.-Q.}\ \bibnamefont
  {Huang}},\ }\href {https://doi.org/10.1103/PhysRevD.83.036005} {\bibfield
  {journal} {\bibinfo  {journal} {Phys. Rev. D}\ }\textbf {\bibinfo {volume}
  {83}},\ \bibinfo {pages} {036005} (\bibinfo {year} {2011})}\BibitemShut
  {NoStop}%
\bibitem [{\citenamefont {Zhao}\ \emph {et~al.}(2025)\citenamefont {Zhao},
  \citenamefont {Kaewsnod}, \citenamefont {Xu}, \citenamefont {Tagsinsit},
  \citenamefont {Liu}, \citenamefont {Limphirat},\ and\ \citenamefont
  {Yan}}]{fourquark5}%
  \BibitemOpen
  \bibfield  {author} {\bibinfo {author} {\bibfnamefont {Z.}~\bibnamefont
  {Zhao}}, \bibinfo {author} {\bibfnamefont {A.}~\bibnamefont {Kaewsnod}},
  \bibinfo {author} {\bibfnamefont {K.}~\bibnamefont {Xu}}, \bibinfo {author}
  {\bibfnamefont {N.}~\bibnamefont {Tagsinsit}}, \bibinfo {author}
  {\bibfnamefont {X.}~\bibnamefont {Liu}}, \bibinfo {author} {\bibfnamefont
  {A.}~\bibnamefont {Limphirat}},\ and\ \bibinfo {author} {\bibfnamefont
  {Y.}~\bibnamefont {Yan}},\ }\href {https://arxiv.org/abs/2503.00552}
  {\bibinfo {title} {Study of $1^{--}$ p wave charmoniumlike and
  bottomoniumlike tetraquark spectroscopy}} (\bibinfo {year} {2025}),\ \Eprint
  {https://arxiv.org/abs/2503.00552} {arXiv:2503.00552 [hep-ph]} \BibitemShut
  {NoStop}%
\bibitem [{\citenamefont {Dubnickova}\ \emph {et~al.}(2020)\citenamefont
  {Dubnickova}, \citenamefont {Dubnicka}, \citenamefont {Issadykov},
  \citenamefont {Ivanov},\ and\ \citenamefont {Liptaj}}]{fourquark8}%
  \BibitemOpen
  \bibfield  {author} {\bibinfo {author} {\bibfnamefont {A.~Z.}\ \bibnamefont
  {Dubnickova}}, \bibinfo {author} {\bibfnamefont {S.}~\bibnamefont
  {Dubnicka}}, \bibinfo {author} {\bibfnamefont {A.}~\bibnamefont {Issadykov}},
  \bibinfo {author} {\bibfnamefont {M.~A.}\ \bibnamefont {Ivanov}},\ and\
  \bibinfo {author} {\bibfnamefont {A.}~\bibnamefont {Liptaj}},\ }\href
  {https://arxiv.org/abs/2003.04142} {\bibinfo {title} {$y(4260)$ as four-quark
  state}} (\bibinfo {year} {2020}),\ \Eprint {https://arxiv.org/abs/2003.04142}
  {arXiv:2003.04142 [hep-ph]} \BibitemShut {NoStop}%
\bibitem [{\citenamefont {Kou}\ and\ \citenamefont {Pene}(2005)}]{hybrid}%
  \BibitemOpen
  \bibfield  {author} {\bibinfo {author} {\bibfnamefont {E.}~\bibnamefont
  {Kou}}\ and\ \bibinfo {author} {\bibfnamefont {O.}~\bibnamefont {Pene}},\
  }\href {https://doi.org/10.1016/j.physletb.2005.09.013} {\bibfield  {journal}
  {\bibinfo  {journal} {Phys. Lett. B}\ }\textbf {\bibinfo {volume} {631}},\
  \bibinfo {pages} {164} (\bibinfo {year} {2005})},\ \Eprint
  {https://arxiv.org/abs/hep-ph/0507119} {arXiv:hep-ph/0507119} \BibitemShut
  {NoStop}%
\bibitem [{\citenamefont {Li}\ and\ \citenamefont {Voloshin}(2014)}]{hy1}%
  \BibitemOpen
  \bibfield  {author} {\bibinfo {author} {\bibfnamefont {X.}~\bibnamefont
  {Li}}\ and\ \bibinfo {author} {\bibfnamefont {M.~B.}\ \bibnamefont
  {Voloshin}},\ }\href {https://doi.org/10.1142/s0217732314500606} {\bibfield
  {journal} {\bibinfo  {journal} {Modern Physics Letters A}\ }\textbf {\bibinfo
  {volume} {29}},\ \bibinfo {pages} {1450060} (\bibinfo {year}
  {2014})}\BibitemShut {NoStop}%
\bibitem [{\citenamefont {Close}\ and\ \citenamefont {Page}(2005)}]{hy3}%
  \BibitemOpen
  \bibfield  {author} {\bibinfo {author} {\bibfnamefont {F.~E.}\ \bibnamefont
  {Close}}\ and\ \bibinfo {author} {\bibfnamefont {P.~R.}\ \bibnamefont
  {Page}},\ }\href
  {https://doi.org/https://doi.org/10.1016/j.physletb.2005.09.016} {\bibfield
  {journal} {\bibinfo  {journal} {Physics Letters B}\ }\textbf {\bibinfo
  {volume} {628}},\ \bibinfo {pages} {215} (\bibinfo {year}
  {2005})}\BibitemShut {NoStop}%
\bibitem [{\citenamefont {Zhu}(2005)}]{hy6}%
  \BibitemOpen
  \bibfield  {author} {\bibinfo {author} {\bibfnamefont {S.-L.}\ \bibnamefont
  {Zhu}},\ }\href
  {https://doi.org/https://doi.org/10.1016/j.physletb.2005.08.068} {\bibfield
  {journal} {\bibinfo  {journal} {Physics Letters B}\ }\textbf {\bibinfo
  {volume} {625}},\ \bibinfo {pages} {212} (\bibinfo {year}
  {2005})}\BibitemShut {NoStop}%
\bibitem [{\citenamefont {Wang}(2016)}]{fourquark6}%
  \BibitemOpen
  \bibfield  {author} {\bibinfo {author} {\bibfnamefont {Z.-G.}\ \bibnamefont
  {Wang}},\ }\bibfield  {journal} {\bibinfo  {journal} {The European Physical
  Journal C}\ }\textbf {\bibinfo {volume} {76}},\ \href
  {https://doi.org/10.1140/epjc/s10052-016-4238-y}
  {10.1140/epjc/s10052-016-4238-y} (\bibinfo {year} {2016})\BibitemShut
  {NoStop}%
\bibitem [{\citenamefont {Wang}(2018{\natexlab{b}})}]{fourquark7}%
  \BibitemOpen
  \bibfield  {author} {\bibinfo {author} {\bibfnamefont {Z.-G.}\ \bibnamefont
  {Wang}},\ }\bibfield  {journal} {\bibinfo  {journal} {The European Physical
  Journal C}\ }\textbf {\bibinfo {volume} {78}},\ \href
  {https://doi.org/10.1140/epjc/s10052-018-5996-5}
  {10.1140/epjc/s10052-018-5996-5} (\bibinfo {year}
  {2018}{\natexlab{b}})\BibitemShut {NoStop}%
\bibitem [{\citenamefont {Qin}\ \emph {et~al.}(2016)\citenamefont {Qin},
  \citenamefont {Xue},\ and\ \citenamefont {Zhao}}]{dd8}%
  \BibitemOpen
  \bibfield  {author} {\bibinfo {author} {\bibfnamefont {W.}~\bibnamefont
  {Qin}}, \bibinfo {author} {\bibfnamefont {S.-R.}\ \bibnamefont {Xue}},\ and\
  \bibinfo {author} {\bibfnamefont {Q.}~\bibnamefont {Zhao}},\ }\href
  {https://doi.org/10.1103/PhysRevD.94.054035} {\bibfield  {journal} {\bibinfo
  {journal} {Phys. Rev. D}\ }\textbf {\bibinfo {volume} {94}},\ \bibinfo
  {pages} {054035} (\bibinfo {year} {2016})}\BibitemShut {NoStop}%
\bibitem [{\citenamefont {Liu}\ \emph {et~al.}(2005)\citenamefont {Liu},
  \citenamefont {Zeng},\ and\ \citenamefont {Li}}]{Liu:2005ay}%
  \BibitemOpen
  \bibfield  {author} {\bibinfo {author} {\bibfnamefont {X.}~\bibnamefont
  {Liu}}, \bibinfo {author} {\bibfnamefont {X.-Q.}\ \bibnamefont {Zeng}},\ and\
  \bibinfo {author} {\bibfnamefont {X.-Q.}\ \bibnamefont {Li}},\ }\href
  {https://doi.org/10.1103/PhysRevD.72.054023} {\bibfield  {journal} {\bibinfo
  {journal} {Phys. Rev. D}\ }\textbf {\bibinfo {volume} {72}},\ \bibinfo
  {pages} {054023} (\bibinfo {year} {2005})},\ \Eprint
  {https://arxiv.org/abs/hep-ph/0507177} {arXiv:hep-ph/0507177} \BibitemShut
  {NoStop}%
\bibitem [{\citenamefont {Yuan}\ \emph {et~al.}(2006)\citenamefont {Yuan},
  \citenamefont {Wang},\ and\ \citenamefont {Mo}}]{dd11}%
  \BibitemOpen
  \bibfield  {author} {\bibinfo {author} {\bibfnamefont {C.}~\bibnamefont
  {Yuan}}, \bibinfo {author} {\bibfnamefont {P.}~\bibnamefont {Wang}},\ and\
  \bibinfo {author} {\bibfnamefont {X.}~\bibnamefont {Mo}},\ }\href
  {https://doi.org/https://doi.org/10.1016/j.physletb.2006.01.031} {\bibfield
  {journal} {\bibinfo  {journal} {Physics Letters B}\ }\textbf {\bibinfo
  {volume} {634}},\ \bibinfo {pages} {399} (\bibinfo {year}
  {2006})}\BibitemShut {NoStop}%
\bibitem [{\citenamefont {van Beveren}\ and\ \citenamefont
  {Rupp}(2008)}]{ccbar4}%
  \BibitemOpen
  \bibfield  {author} {\bibinfo {author} {\bibfnamefont {E.}~\bibnamefont {van
  Beveren}}\ and\ \bibinfo {author} {\bibfnamefont {G.}~\bibnamefont {Rupp}},\
  }\href {https://arxiv.org/abs/0811.1755} {\bibinfo {title} {The spectrum of
  charmonium in the resonance-spectrum expansion}} (\bibinfo {year} {2008}),\
  \Eprint {https://arxiv.org/abs/0811.1755} {arXiv:0811.1755 [hep-ph]}
  \BibitemShut {NoStop}%
\bibitem [{\citenamefont {Chen}\ \emph {et~al.}(2016)\citenamefont {Chen},
  \citenamefont {Liu}, \citenamefont {Li},\ and\ \citenamefont {Ke}}]{ccbar5}%
  \BibitemOpen
  \bibfield  {author} {\bibinfo {author} {\bibfnamefont {D.-Y.}\ \bibnamefont
  {Chen}}, \bibinfo {author} {\bibfnamefont {X.}~\bibnamefont {Liu}}, \bibinfo
  {author} {\bibfnamefont {X.-Q.}\ \bibnamefont {Li}},\ and\ \bibinfo {author}
  {\bibfnamefont {H.-W.}\ \bibnamefont {Ke}},\ }\href
  {https://doi.org/10.1103/PhysRevD.93.014011} {\bibfield  {journal} {\bibinfo
  {journal} {Phys. Rev. D}\ }\textbf {\bibinfo {volume} {93}},\ \bibinfo
  {pages} {014011} (\bibinfo {year} {2016})}\BibitemShut {NoStop}%
\bibitem [{\citenamefont {Bukin}(2007)}]{Bukin:2007kx}%
  \BibitemOpen
  \bibfield  {author} {\bibinfo {author} {\bibfnamefont {A.~D.}\ \bibnamefont
  {Bukin}},\ }\href@noop {} {\  (\bibinfo {year} {2007})},\ \Eprint
  {https://arxiv.org/abs/0710.5627} {arXiv:0710.5627 [physics.data-an]}
  \BibitemShut {NoStop}%
\bibitem [{\citenamefont {Zhu}\ \emph {et~al.}(2011)\citenamefont {Zhu},
  \citenamefont {Mo}, \citenamefont {Yuan},\ and\ \citenamefont
  {Wang}}]{Zhu:2011ha}%
  \BibitemOpen
  \bibfield  {author} {\bibinfo {author} {\bibfnamefont {K.}~\bibnamefont
  {Zhu}}, \bibinfo {author} {\bibfnamefont {X.~H.}\ \bibnamefont {Mo}},
  \bibinfo {author} {\bibfnamefont {C.~Z.}\ \bibnamefont {Yuan}},\ and\
  \bibinfo {author} {\bibfnamefont {P.}~\bibnamefont {Wang}},\ }\href
  {https://doi.org/10.1142/S0217751X11054589} {\bibfield  {journal} {\bibinfo
  {journal} {Int. J. Mod. Phys. A}\ }\textbf {\bibinfo {volume} {26}},\
  \bibinfo {pages} {4511} (\bibinfo {year} {2011})},\ \Eprint
  {https://arxiv.org/abs/1108.2760} {arXiv:1108.2760 [hep-ex]} \BibitemShut
  {NoStop}%
\bibitem [{\citenamefont {Bai}\ and\ \citenamefont {Chen}(2019)}]{Bai:2019jrb}%
  \BibitemOpen
  \bibfield  {author} {\bibinfo {author} {\bibfnamefont {Y.}~\bibnamefont
  {Bai}}\ and\ \bibinfo {author} {\bibfnamefont {D.-Y.}\ \bibnamefont {Chen}},\
  }\href {https://doi.org/10.1103/PhysRevD.99.072007} {\bibfield  {journal}
  {\bibinfo  {journal} {Phys. Rev. D}\ }\textbf {\bibinfo {volume} {99}},\
  \bibinfo {pages} {072007} (\bibinfo {year} {2019})},\ \Eprint
  {https://arxiv.org/abs/1901.01394} {arXiv:1901.01394 [hep-ph]} \BibitemShut
  {NoStop}%
\bibitem [{\citenamefont {Ablikim}\ \emph
  {et~al.}(2022{\natexlab{a}})\citenamefont {Ablikim} \emph
  {et~al.}}]{pipijpsi}%
  \BibitemOpen
  \bibfield  {author} {\bibinfo {author} {\bibfnamefont {M.}~\bibnamefont
  {Ablikim}} \emph {et~al.} (\bibinfo {collaboration} {BESIII Collaboration}),\
  }\href {https://doi.org/10.1103/PhysRevD.106.072001} {\bibfield  {journal}
  {\bibinfo  {journal} {Phys. Rev. D}\ }\textbf {\bibinfo {volume} {106}},\
  \bibinfo {pages} {072001} (\bibinfo {year} {2022}{\natexlab{a}})}\BibitemShut
  {NoStop}%
\bibitem [{\citenamefont {Ablikim}\ \emph
  {et~al.}(2020{\natexlab{a}})\citenamefont {Ablikim} \emph
  {et~al.}}]{pi0pi0jpsi}%
  \BibitemOpen
  \bibfield  {author} {\bibinfo {author} {\bibfnamefont {M.}~\bibnamefont
  {Ablikim}} \emph {et~al.} (\bibinfo {collaboration} {BESIII}),\ }\href
  {https://doi.org/10.1103/PhysRevD.102.012009} {\bibfield  {journal} {\bibinfo
   {journal} {Phys. Rev. D}\ }\textbf {\bibinfo {volume} {102}},\ \bibinfo
  {pages} {012009} (\bibinfo {year} {2020}{\natexlab{a}})},\ \Eprint
  {https://arxiv.org/abs/2004.13788} {arXiv:2004.13788 [hep-ex]} \BibitemShut
  {NoStop}%
\bibitem [{\citenamefont {Ablikim}\ \emph
  {et~al.}(2022{\natexlab{b}})\citenamefont {Ablikim} \emph {et~al.}}]{kkjpsi}%
  \BibitemOpen
  \bibfield  {author} {\bibinfo {author} {\bibfnamefont {M.}~\bibnamefont
  {Ablikim}} \emph {et~al.},\ }\href {https://doi.org/10.1088/1674-1137/ac945c}
  {\bibfield  {journal} {\bibinfo  {journal} {Chinese Physics C}\ }\textbf
  {\bibinfo {volume} {46}},\ \bibinfo {pages} {111002} (\bibinfo {year}
  {2022}{\natexlab{b}})}\BibitemShut {NoStop}%
\bibitem [{\citenamefont {Ablikim}\ \emph {et~al.}(2023)\citenamefont {Ablikim}
  \emph {et~al.}}]{ksksjpsi}%
  \BibitemOpen
  \bibfield  {author} {\bibinfo {author} {\bibfnamefont {M.}~\bibnamefont
  {Ablikim}} \emph {et~al.} (\bibinfo {collaboration} {BESIII Collaboration}),\
  }\href {https://doi.org/10.1103/PhysRevD.107.092005} {\bibfield  {journal}
  {\bibinfo  {journal} {Phys. Rev. D}\ }\textbf {\bibinfo {volume} {107}},\
  \bibinfo {pages} {092005} (\bibinfo {year} {2023})}\BibitemShut {NoStop}%
\bibitem [{\citenamefont {Ablikim}\ \emph {et~al.}(2024)\citenamefont {Ablikim}
  \emph {et~al.}}]{etajpsi}%
  \BibitemOpen
  \bibfield  {author} {\bibinfo {author} {\bibfnamefont {M.}~\bibnamefont
  {Ablikim}} \emph {et~al.} (\bibinfo {collaboration} {BESIII}),\ }\href
  {https://doi.org/10.1103/PhysRevD.109.092012} {\bibfield  {journal} {\bibinfo
   {journal} {Phys. Rev. D}\ }\textbf {\bibinfo {volume} {109}},\ \bibinfo
  {pages} {092012} (\bibinfo {year} {2024})},\ \Eprint
  {https://arxiv.org/abs/2310.03361} {arXiv:2310.03361 [hep-ex]} \BibitemShut
  {NoStop}%
\bibitem [{\citenamefont {Ablikim}\ \emph
  {et~al.}(2020{\natexlab{b}})\citenamefont {Ablikim} \emph
  {et~al.}}]{eta'jpsi}%
  \BibitemOpen
  \bibfield  {author} {\bibinfo {author} {\bibfnamefont {M.}~\bibnamefont
  {Ablikim}} \emph {et~al.} (\bibinfo {collaboration} {BESIII Collaboration}),\
  }\href {https://doi.org/10.1103/PhysRevD.101.012008} {\bibfield  {journal}
  {\bibinfo  {journal} {Phys. Rev. D}\ }\textbf {\bibinfo {volume} {101}},\
  \bibinfo {pages} {012008} (\bibinfo {year} {2020}{\natexlab{b}})}\BibitemShut
  {NoStop}%
\bibitem [{\citenamefont {Ablikim}\ \emph {et~al.}(2017)\citenamefont {Ablikim}
  \emph {et~al.}}]{pipipsip}%
  \BibitemOpen
  \bibfield  {author} {\bibinfo {author} {\bibfnamefont {M.}~\bibnamefont
  {Ablikim}} \emph {et~al.} (\bibinfo {collaboration} {BESIII Collaboration}),\
  }\href {https://doi.org/10.1103/PhysRevD.96.032004} {\bibfield  {journal}
  {\bibinfo  {journal} {Phys. Rev. D}\ }\textbf {\bibinfo {volume} {96}},\
  \bibinfo {pages} {032004} (\bibinfo {year} {2017})}\BibitemShut {NoStop}%
\bibitem [{\citenamefont {Zhu}(2022)}]{eta-eta'}%
  \BibitemOpen
  \bibfield  {author} {\bibinfo {author} {\bibfnamefont {K.}~\bibnamefont
  {Zhu}},\ }\href {https://doi.org/10.1103/PhysRevD.105.L031506} {\bibfield
  {journal} {\bibinfo  {journal} {Phys. Rev. D}\ }\textbf {\bibinfo {volume}
  {105}},\ \bibinfo {pages} {L031506} (\bibinfo {year} {2022})}\BibitemShut
  {NoStop}%
\bibitem [{\citenamefont {Qiao}(2006)}]{explanation}%
  \BibitemOpen
  \bibfield  {author} {\bibinfo {author} {\bibfnamefont {C.-F.}\ \bibnamefont
  {Qiao}},\ }\href
  {https://doi.org/https://doi.org/10.1016/j.physletb.2006.06.038} {\bibfield
  {journal} {\bibinfo  {journal} {Physics Letters B}\ }\textbf {\bibinfo
  {volume} {639}},\ \bibinfo {pages} {263} (\bibinfo {year}
  {2006})}\BibitemShut {NoStop}%
\bibitem [{\citenamefont {Navas}\ \emph {et~al.}(2024)\citenamefont {Navas}
  \emph {et~al.}}]{pdg}%
  \BibitemOpen
  \bibfield  {author} {\bibinfo {author} {\bibfnamefont {S.}~\bibnamefont
  {Navas}} \emph {et~al.} (\bibinfo {collaboration} {Particle Data Group}),\
  }\href {https://doi.org/10.1103/PhysRevD.110.030001} {\bibfield  {journal}
  {\bibinfo  {journal} {Phys. Rev. D}\ }\textbf {\bibinfo {volume} {110}},\
  \bibinfo {pages} {030001} (\bibinfo {year} {2024})}\BibitemShut {NoStop}%
\bibitem [{\citenamefont {Ganbold}(2024)}]{hiddencharm}%
  \BibitemOpen
  \bibfield  {author} {\bibinfo {author} {\bibfnamefont {G.}~\bibnamefont
  {Ganbold}},\ }\href {https://doi.org/10.1134/S1063779624700175} {\bibfield
  {journal} {\bibinfo  {journal} {Phys. Part. Nucl.}\ }\textbf {\bibinfo
  {volume} {55}},\ \bibinfo {pages} {781} (\bibinfo {year} {2024})}\BibitemShut
  {NoStop}%
\bibitem [{\citenamefont {Ablikim}\ \emph {et~al.}(2025)\citenamefont {Ablikim}
  \emph {et~al.}}]{BESIIIPWA}%
  \BibitemOpen
  \bibfield  {author} {\bibinfo {author} {\bibfnamefont {M.}~\bibnamefont
  {Ablikim}} \emph {et~al.} (\bibinfo {collaboration} {BESIII}),\ }\href@noop
  {} {\  (\bibinfo {year} {2025})},\ \Eprint {https://arxiv.org/abs/2505.13222}
  {arXiv:2505.13222 [hep-ex]} \BibitemShut {NoStop}%
\end{thebibliography}%
\end{document}